\documentstyle[epsf,12pt]{article}
\begin{document}

%%%%%%%%%%%%%%%%%%%%%%%%%%%%%%%%%%%%%%%%%%%%%%%%%%%%%%%%%%%%%%
\catcode`@=11
% Redefine caption to put text and formulas in smaller font
\long\def\@caption#1[#2]#3{\par\addcontentsline{\csname
  ext@#1\endcsname}{#1}{\protect\numberline{\csname
  the#1\endcsname}{\ignorespaces #2}}\begingroup
    \small
    \@parboxrestore
    \@makecaption{\csname fnum@#1\endcsname}{\ignorespaces #3}\par
  \endgroup}
\catcode`@=12
%%%%%%%%%%%%%%%%%%%%%%%%%%%%%%%%%%%%%%%%%%%%%%%%%%%%%%%%%%%%
\newcommand{\newc}{\newcommand}
\newc{\gsim}{\lower.7ex\hbox{$\;\stackrel{\textstyle>}{\sim}\;$}}
\newc{\lsim}{\lower.7ex\hbox{$\;\stackrel{\textstyle<}{\sim}\;$}}
\newc{\gev}{\,{\rm GeV}}
\newc{\mev}{\,{\rm MeV}}
\newc{\ev}{\,{\rm eV}}
\newc{\kev}{\,{\rm keV}}
\newc{\tev}{\,{\rm TeV}}
\newc{\mz}{m_Z}
\newc{\mpl}{M_{Pl}}
\newc{\chifc}{\chi_{{}_{\!F\!C}}}
\newc\order{{\cal O}}
\newc\CO{\order}
\newc\CL{{\cal L}}
\newc\CY{{\cal Y}}
\newc\CH{{\cal H}}
\newc\CM{{\cal M}}
\newc\CF{{\cal F}}
\newc\CD{{\cal D}}
\newc\CN{{\cal N}}
\newc{\eps}{\epsilon}
\newc{\re}{\mbox{Re}\,}
\newc{\im}{\mbox{Im}\,}
\newc{\invpb}{\,\mbox{pb}^{-1}}
\newc{\invfb}{\,\mbox{fb}^{-1}}
\newc{\yddiag}{{\bf D}}
\newc{\yddiagd}{{\bf D^\dagger}}
\newc{\yudiag}{{\bf U}}
\newc{\yudiagd}{{\bf U^\dagger}}
\newc{\yd}{{\bf Y_D}}
\newc{\ydd}{{\bf Y_D^\dagger}}
\newc{\yu}{{\bf Y_U}}
\newc{\yud}{{\bf Y_U^\dagger}}
\newc{\ckm}{{\bf V}}
\newc{\ckmd}{{\bf V^\dagger}}
\newc{\ckmz}{{\bf V^0}}
\newc{\ckmzd}{{\bf V^{0\dagger}}}
\newc{\X}{{\bf X}}
\newc{\bbbar}{B^0-\bar B^0}
\def\bra#1{\left\langle #1 \right|}
\def\ket#1{\left| #1 \right\rangle}
\newc{\sgn}{\mbox{sgn}\,}
\newc{\m}{{\bf m}}
\newc{\msusy}{M_{\rm SUSY}}
\newc{\munif}{M_{\rm unif}}
\newc{\slepton}{{\tilde\ell}}
\newc{\Slepton}{{\tilde L}}
\newc{\sneutrino}{{\tilde\nu}}
\newc{\selectron}{{\tilde e}}
\newc{\stau}{{\tilde\tau}}
%
%%%%%%%%%%%%%%%%%% Reference Defs %%%%%%%%%%%%%%%%%%
%
\def\NPB#1#2#3{Nucl. Phys. {\bf B#1} (19#2) #3}
\def\PLB#1#2#3{Phys. Lett. {\bf B#1} (19#2) #3}
\def\PLBold#1#2#3{Phys. Lett. {\bf#1B} (19#2) #3}
\def\PRD#1#2#3{Phys. Rev. {\bf D#1} (19#2) #3}
\def\PRL#1#2#3{Phys. Rev. Lett. {\bf#1} (19#2) #3}
\def\PRT#1#2#3{Phys. Rep. {\bf#1} (19#2) #3}
\def\ARAA#1#2#3{Ann. Rev. Astron. Astrophys. {\bf#1} (19#2) #3}
\def\ARNP#1#2#3{Ann. Rev. Nucl. Part. Sci. {\bf#1} (19#2) #3}
\def\MPL#1#2#3{Mod. Phys. Lett. {\bf #1} (19#2) #3}
\def\ZPC#1#2#3{Zeit. f\"ur Physik {\bf C#1} (19#2) #3}
\def\APJ#1#2#3{Ap. J. {\bf #1} (19#2) #3}
\def\AP#1#2#3{{Ann. Phys. } {\bf #1} (19#2) #3}
\def\RMP#1#2#3{{Rev. Mod. Phys. } {\bf #1} (19#2) #3}
\def\CMP#1#2#3{{Comm. Math. Phys. } {\bf #1} (19#2) #3}
\relax
%
%
%%%%%%%%%%%%%%%%%%%%%%% latex eqn abrev's %%%%%%%%%%%%%%%%%%%%%%%%%%%%
%
\def\beq{\begin{equation}}
\def\eeq{\end{equation}}
\def\bea{\begin{eqnarray}}
\def\eea{\end{eqnarray}}
%
%%%%%%%%%%%%%%%%%%%%%%% common abrev's %%%%%%%%%%%%%%%%%
%
%
\newc{\ie}{{\it i.e.}}          \newc{\etal}{{\it et al.}}
\newc{\eg}{{\it e.g.}}          \newc{\etc}{{\it etc.}}
\newc{\cf}{{\it c.f.}}
\def\smuon{{\tilde\mu}}
\def\neut{{\tilde N}}
\def\char{{\tilde C}}
\def\bino{{\tilde B}}
\def\wino{{\tilde W}}
\def\higgsino{{\tilde H}}
\def\sneut{{\tilde\nu}}
%
%
%%%%%%%%%%%%%%%%%%%% slashed symbols %%%%%%%%%%%%%%%%%%%%%
%
%
\def\slash#1{\rlap{$#1$}/} % slashes a character
\def\Dsl{\,\raise.15ex\hbox{/}\mkern-13.5mu D} %this one can be subscripted
\def\delsl{\raise.15ex\hbox{/}\kern-.57em\partial}
\def\Ksl{\hbox{/\kern-.6000em\rm K}}
\def\Asl{\hbox{/\kern-.6500em \rm A}}
\def\Qsl{\hbox{/\kern-.6000em\rm Q}}
\def\gradsl{\hbox{/\kern-.6500em$\nabla$}}
%
%%%%%%%%%%%%%%%%%%% various symbol abbreviations, vev's etc %%%%%%%%%%%
%
%
\def\bar#1{\overline{#1}}
\def\vev#1{\left\langle #1 \right\rangle}
%
%%%%%%%%%%%%%%%%%%% end of intro %%%%%%%%%%%%%%%%%%%%%%%%%%%%%%%%%%%%%

\begin{titlepage} 
\begin{flushright}
FCFM-BUAP 
\end{flushright}
\vskip 2cm
\begin{center}
{\large\bf Flavor Symmetries in Extra Dimensions}
\vskip 1cm
{\normalsize\bf

Alfredo Aranda$^a$ and J. Lorenzo Diaz-Cruz$^b$ \\
\vskip 0.5cm
$^{(a)}${\it 
Facultad de Ciencias, Universidad de Colima\\
Colima, Col., M\'exico}\\ 
$^{(b)}${\it Center for Particles and Fields, FCFM-BUAP\\
Puebla, Pue., M\'exico}\\ 
~~ \\
}

\end{center}
\vskip .5cm

\begin{abstract}
We present a class of  models of flavor that rely on the
use of flavor symmetries and the Froggart-Nielsen mechanism 
in extra dimensions.  
The particle content is that of the standard model plus an 
additional flavon field; all the fields  propagate in universal 
extra dimensions and the flavor scale is associated with 
the cutoff of the theory, which in 5D is $\sim 10$~TeV. 
The Yukawa matrices are generated by higher dimension operators involving
flavon fields, whose vacuum expectation values break the flavor 
symmetry. We apply this framework and present a specific 5D model 
based on a discrete local symmetry that reproduces all fermion masses 
and mixing angles both in the quark and charged lepton sectors. 
\end{abstract}

\end{titlepage}

\setcounter{footnote}{0}
\setcounter{page}{1}
\setcounter{section}{0}
\setcounter{subsection}{0}
\setcounter{subsubsection}{0}

%%%%%%%%%%%%%%%%%%%%%%%%%%%%%%%%%%%%%%%%%%%%%%%%%%%%%%%%%%%%%%%%%%%%%%%

\section{Introduction} \label{sec:intro}
It is possible that there exist extra dimensions and that they might
even play an important role in electroweak physics. This possibility has led
to an impressive amount of work over the past few years resulting in 
new and exciting ways of tackling (or interpreting) some of 
the main problems in particle physics. Models with one or more
extra dimensions, flat or warped, with and without SUSY have been
discussed in the literature. For a partial list of references see~\cite{XD}.

The problem of flavor is one of the pressing problems in particle physics.
In the context of extra dimensions some interesting solutions have been 
presented. For example it is possible to generate hierarchies among the 
fermion masses and mixing 
parameters by restricting them to a brane while imposing some
flavor symmetry broken by the mechanism of shinning~\cite{nima1}. 
Another way is to localize the
fermion fields in different positions along the extra dimensions, 
and in this way
generate hierarchies among masses and mixing parameters through wave-function
overlaps~\cite{martin1}. Yet another alternative, similar in 
spirit to the previous ``cartographic'' solution, is to localize the fermion
fields along different points inside a fat brane~\cite{fat} or by placing 
different matter fields 
in different branes~\cite{dvali1}. In the fat-brane case, the hierarchies 
and mixing properties are
obtained, as before, through the wave-function overlaps. In the case of 
several branes this is
accomplished by having the branes intersect in some specific way. 
Other new mechanisms employ warped extra dimensions~\cite{randall} to, 
for example, produce small Yukawa couplings and hierarchies, 
as well as small neutrino masses without the need of a seesaw~\cite{flavor}.

However, despite the new ideas brought by extra dimensions to the flavor 
problem, it does not seem to be possible to generate realistic masses 
within the minimal frameworks proposed thus far, and one may need to invoke
a certain amount of flavor symmetries to solve the problem. In this letter 
we discuss a class of extra-dimensional models of flavor, with the following 
properties: 

\begin{itemize}
\item Spacetime consists of $4+\delta$-dimensions in which all fields 
propagate, i.e. they are universal. The extra dimensions are compactified 
on generalized orbifolds $T^n/Z_2$; which in 5D have a radius of 
compactification  $R$.
\item A flavor symmetry is added to the SM gauge group. 
This symmetry is broken by the vacuum expectation values (vevs) 
of a flavon field, leading to the generation of Yukawa matrix
textures for all matter fields through the Froggart-Nielsen (FN)
mechanism. 
\item The particle content of the model is that of the SM plus 
one additional scalar (flavon) field. This is the minimal set 
we require in order to reproduce the observed hierarchies.
\end{itemize}

We show that given the above conditions, it is possible to generate viable 
models of flavor that reproduces all observed masses and mixing angles, 
both in the quark and charged lepton sectors. Furthermore, this is 
accomplished  with a flavor scale determined by the current perturbativity 
bound on the scale of the universal extra dimension, which for 5D is 
of order $\sim 10$~TeV~\cite{dobrescu1}. The properties of the model are 
chosen so as to minimize the amount of additions beyond the SM.
It should be noted that the flavor physics presented 
in this paper is based on the traditional Froggart-Nielsen 
mechanism with flavor symmetries, and the addition of extra dimensions 
is done in order to obtain a low flavor scale.

The framework of our extra-dimensional flavor models is presented in 
section \ref{sec:fnexdm}. There we show how to generate the Yukawa 
matrices from operators in the extra-dimensional Lagrangian using the FN 
mechanism. In Section~\ref{sec:results}, we present an 
specific model in 5 dimensions, and show the results from a fit 
to the experimental data. Additional comments on the model and
its extension to 6 dimensions are presented in Sect. \ref{sec:sixdim}, 
which also contains our conclusions.

\section{The FN mechanism in extra-dimensions} \label{sec:fnexdm}

As mentioned in the introduction, the model consists of the SM fields plus 
one flavon field, all propagating in a $(4+\delta)$-dimensional spacetime. 
We assume the extra-dimensional coordinates ($y_i$, $i=1,..\delta$) 
are compactified on a $\delta$-torus/$Z_2$; in 5d it corresponds to an
orbifold $S^1/Z_2$ with a radius of compactification $R = 1/M_c$, 
where $M_c$ is the compactification scale. 

In $(4+\delta)$-dimensions, the Lagrangian has mass-dimensions
$4+\delta$; therefore the fermion fields have dimensions
$(3+\delta)/2$, while for the scalar fields it is: $(2+\delta)/2$
and the XD Yukawa coupling have dimensions $-\delta/2$.
 
Thus, the fermion fields (quarks and leptons) can be decomposed 
in $(4+\delta)$-dimensions as follows~\cite{dobrescu1}

\begin{eqnarray} \label{decomposition-fermions}
{\cal{Q}}(x^{\mu},y)  = 
   \frac{1}{(\pi R)^{\delta/2}} \left( Q^{(0)}_L(x^{\mu})
+ \sqrt{2} \sum_{[n_i]}^{\infty}\left[ P_L Q_L^{(n_i)}(x^{\mu}) f^{n_i}_L(y_i)
+ P_R Q_R^{(n_i)}(x^{\mu}) f^{n_i}_R(y_i) \right] \right) \, , \\ \nonumber \\ 
{\cal{E}}(x^{\mu},y) = \frac{1}{(\pi R)^{\delta/2}} \left( E^{(0)}_R(x^{\mu})
+ \sqrt{2} \sum_{[n_i]}^{\infty}\left[ P_R E_R^{(n_i)}(x^{\mu})f^{n_i}_L(y_i)  
+ P_L E_L^{(n_i)}(x^{\mu}) f^{n_i}_R(y_i) \right] \right) \, ,
\end{eqnarray}
where ${\cal Q}$ and ${\cal E}$ denote an ${\rm SU(2)}_W$ doublet and 
singlet field respectively. $P_L$ and $P_R$ are the 4D chiral projection 
operators $P_{R,L} = (1 \pm \gamma_5)/2$. The sum is performed over 
the complete set of indexes $[n_i]$ needed to specify the KK expansion. 
This decomposition should ensure that the zero modes correspond to SM fields. 
For instance, in 5-dimensions the expansion functions correspond to:  
$f^{n_i}_L=\cos\left(\frac{ny}{R}\right)$,
$f^{n_i}_R=\sin\left(\frac{ny}{R}\right)$,
while their generalizations to six-dimensions can be found in 
Ref. ~\cite{dobrescu1} too.

Since the Higgs and the flavon fields 
must couple to these zero-modes, they must be even under the $Z_2$.
Their decomposition is then given by
\begin{eqnarray} \label{decomposition-scalar}
{\cal{S}}(x^{\mu},y) = \frac{1}{(\pi R)^{\delta/2}} 
\left( S^{(0)}(x^{\mu})
+ \sqrt{2} \sum_{[n_i]}^{\infty} S^{(n_i)}(x^{\mu}) f^{n_i}_L(y_i) \right) \, .
\end{eqnarray}

Now we can consider the generation of Yukawa matrices.
Once a flavor symmetry is included in the model, the lowest dimensional
Yukawa terms are absent, though they can appear as higher-order
operators. All the terms in the Lagrangian responsible for the fermion 
(modulo neutrinos) masses will in general have the form:

\begin{eqnarray} \label{general}
{\cal L}_{XD} \sim \hat{\lambda}_{ab} \bar{{\cal Q}}_{a} 
         i\sigma_{2}{\cal H}^*{\cal U}_{b} 
[\frac{\Phi}{\Lambda^{(2+\delta)/2}}]^n + h.c. \, ,
\end{eqnarray}
where $a$ and $b$ are flavor indexes. Dimensionless couplings are
introduced by defining:
$\hat{\lambda}_{ab} = \lambda_{ab}[\pi R]^{\delta/2}$, 
$\lambda_{ab}$ denotes the 4D Yukawa coupling
(which we assume to be a number of order one), 
$\Lambda$ is the cutoff of the theory, and $\Phi$ represents a flavon field. 
 To see how the 4D Yukawa matrices are generated, lets consider 
Eq.~(\ref{general}) after compactification. We are assuming
that the flavor symmetry is broken by the vevs of the flavon field at or 
very close to the scale $\Lambda$. Thus, after compactification, we obtain

\begin{eqnarray} \label{general4D}
{\cal L}^4 \sim \lambda_{ab} \bar{Q}_a i \sigma_2 H^*  U_b \left[
\left(\frac{M_c}{\pi \Lambda}\right)^{n \delta/2}\right] + h.c. \, ,
\end{eqnarray}
where all fields now correspond to the zero-modes of those in 
Eq.~(\ref{general}), and where the vev of $\Phi$ has been set 
equal to $\Lambda$. We now define a new 4D Yukawa coupling given by

\begin{eqnarray} \label{yukawa4D}
\lambda_{ab}^{\prime} = 
\lambda_{ab}\left(\frac{M_c}{\pi \Lambda}\right)^{n\delta/2} =
{\rm O(1)} \left(\frac{M_c}{\pi \Lambda}\right)^{n\delta/2} = 
{\rm O(1)} \epsilon^{n\delta} \, ,
\end{eqnarray}
where  $\lambda_{ab} = $~O(1) and $\epsilon=\sqrt{M_c/\pi \Lambda}$
should be of order 0.1. Thus, the hierarchies in masses and mixing 
angles are obtained by assigning different charges to different 
generations in such a way as to obtain realistic textures for the 
Yukawa matrices of quarks and leptons. These matrices appear as an 
expansion in the small parameter $\epsilon$.
 
\section{A 5D Model} \label{sec:results}

We are now in a position to present a flavor model in 5 dimensions. 
Using the values for $M_c = 0.3$~TeV, and $\Lambda = 10$~TeV from 
Ref.~\cite{dobrescu1}, we obtain that the expansion parameter
is $\epsilon \approx 0.1$. The 5D model is based on a $Z_{6}$ 
local flavor symmetry whose anomalies are assumed to be canceled by a 
Green-Schwarz mechanism~\cite{green} (For a discussion
on discrete gauge symmetries see~\cite{discrete}). 

The charge assignments for the matter fields are given by 
\begin{eqnarray} \label{charges}
\nonumber
{\cal Q} \sim (0,5,4) \,\, , {\bar{{\cal Q}}} \sim (0,1,2) \\ \nonumber
{\cal L} \sim (0,5,4) \,\, , {\bar{{\cal L}}} \sim (0,1,2) \\ \nonumber
{\cal U} \sim (2,3,4) \,\, , \,\, {\cal D} \sim (1,2,2) \,\, , \,\,
{\cal E} \sim (1,2,2) \, ,
\end{eqnarray}
where the numbers in parenthesis correspond to the charges of each
generation. 

The charges for the
Higgs (${\cal H}$) and flavon field ($\Phi$) are 
\begin{eqnarray} \label{chargesscalars}
{\cal H} \sim 0 \,\, , \,\, \Phi \sim 1 
\end{eqnarray}
Using these assignments together with Eqs.~(\ref{general4D}) 
and~(\ref{yukawa4D}) to compute the
Yukawa matrices we obtain
\begin{eqnarray} \label{yukawau}
\lambda_U \sim 
\left(
\begin{array}{ccc}
\phi^4 & \phi^3 & \phi^2 \\
\phi^3 & \phi^2 & \phi \\
\phi^2 & \phi & 1
\end{array}
\right) \rightarrow
\left(
\begin{array}{ccc}
\epsilon^4 & \epsilon^3 & \epsilon^2 \\
\epsilon^3 & \epsilon^2 & \epsilon \\
\epsilon^2 & \epsilon & 1 
\end{array}
\right) \,\, , \\ \nonumber \\
\lambda_D \sim 
\left(
\begin{array}{ccc}
\phi^5 & \phi^4 & \phi^4 \\
\phi^4 & \phi^3 & \phi^3 \\
\phi^3 & \phi^2 & \phi^2
\end{array}
\right) \rightarrow
\left(
\begin{array}{ccc}
\epsilon^5 & \epsilon^4 & \epsilon^4 \\
\epsilon^4 & \epsilon^3 & \epsilon^3 \\
\epsilon^3 & \epsilon^2 & \epsilon^2 
\end{array}
\right) \,\, , \\ \nonumber \\
\lambda_E \sim 
\left(
\begin{array}{ccc}
\phi^5 & \phi^4 & \phi^4 \\
\phi^4 & \phi^3 & \phi^3 \\
\phi^3 & \phi^2 & \phi^2
\end{array}
\right) \rightarrow
\left(
\begin{array}{ccc}
\epsilon^5 & \epsilon^4 & \epsilon^4 \\
\epsilon^4 & \epsilon^3 & \epsilon^3 \\
\epsilon^3 & \epsilon^2 & \epsilon^2 
\end{array}
\right) \,\, , 
\end{eqnarray}
where O(1) coefficients have been omitted, and only the powers of $\phi$ are 
shown here for clarity; $\phi$ represents the zero-mode of $\Phi$. 

On the other hand, a possible way to compute the
neutrino mass matrix, is through the following operator
\begin{eqnarray} \label{neutrino5D}
{\cal L}^5 \sim l_{ab} \bar{{\cal L}}^c_a {\cal L}_b {\cal H}^2 \frac{\Phi^n}{\Lambda^{3n/2+2}} \, 
\end{eqnarray}
where $l_{ab}$ is an O(1) parameter. 
After compactification this operator becomes
indeed a mass matrix operator, namely
\begin{eqnarray} \label{neutrino4D}
{\cal L}^4 \sim l_{ab}^{\prime} \bar{L}^c_a L_b H^2 \, ,
\end{eqnarray}
where $l_{ab}^{\prime} = {\rm O(1)}\frac{\epsilon^{(n+2)}}{\Lambda}$,
is a Yukawa parameter. Using these operators one can generate textures
that reproduce the mixings and mass hierarchies in this sector, but
in order to get the right neutrino scale,
one might need, for example, to introduce other flavons propagating in the bulk of
more than one extra dimension. Alternatively one could
introduce right-handed neutrinos $\nu_R$ propagating in more extra dimensions.
In this work, we will not pursue further the problem of
neutrino masses. We just mention that
it is indeed possible to construct a model that does the job \cite{progress}
and concentrate in presenting a detailed 
discussion on the mass matrices for the quark and charged lepton 
sectors.

Let us now discuss how the textures obtained above reproduce 
the observed mass patterns and mixing angles in both the quark and charged lepton 
sectors. To do this, we need to include O(1) coefficients in the entries of 
the Yukawa matrices. These coefficients are 
determined by performing a fit to the observables.
We emphasize that the hierarchies among the masses and mixing angles 
are determined by
the textures and not by the coefficients. In order
to determine that this is the case, we perform several fits starting 
from randomly selected
initial sets of parameters. An important property of the fits is 
that the O(1) parameters are 
not treated freely and they are
allowed to vary only within a range, say between $1/3$ and $3$. 
They are then 
included into a $\chi^2$
function and thus treated as additional pieces of data. 
The particular range is of course arbitrary
and is meant only to determine what we explicitly mean by O(1). 
For details about the fit and
how to treat the O(1) coefficients see Ref.~\cite{aranda}.

In the quark sector we fit to the six quark masses and three CKM-angles 
(CP violation is neglected)
whereas in the charged lepton sector we use the $e$, $\mu$, and $\tau$ 
masses.
The experimental uncertainties on the observables (or estimates in the 
case of the quark masses)
used in the fits are either those of Ref~\cite{PDG} or $1\%$ 
of the central value, whichever is larger.
We find very good fits for a large number of initial points and conclude that the textures 
presented in the previous section do reproduce the observed patterns. 
In Tables~\ref{fit} and~\ref{data} we present the results of one of the fits where we used
the following parametrizations:
\begin{eqnarray} \label{parametrization}
\lambda_U = \left(
\begin{array}{ccc}
u_1 \epsilon^4 & u_2 \epsilon^3 & u_3 \epsilon^2 \\
u_4 \epsilon^3 & u_5 \epsilon^2 & u_6 \epsilon \\
u_7 \epsilon^2 & u_8 \epsilon & u_9  
\end{array}
\right) \,\, , \, \,
\lambda_D = \left(
\begin{array}{ccc}
d_1 \epsilon^5 & d_2 \epsilon^4 & d_3 \epsilon^4 \\
d_4 \epsilon^4 & d_5 \epsilon^3 & d_6 \epsilon^3 \\
d_7 \epsilon^3 & d_8 \epsilon^2 & d_9 \epsilon^2 
\end{array}
\right) \,\, , \, \,
\lambda_E = \left(
\begin{array}{ccc}
l_1 \epsilon^5 & l_2 \epsilon^4 & l_3 \epsilon^4 \\
l_4 \epsilon^4 & l_5 \epsilon^3 & l_6 \epsilon^3 \\
l_7 \epsilon^3 & l_8 \epsilon^2 & l_9 \epsilon^2 
\end{array}
\right) .
\end{eqnarray}

We will present a complete analysis involving a large number of fits, 
including the neutrino sector as well, in a
longer version of this letter~\cite{progress}.

\section{Comments and Conclusions} \label{sec:sixdim}

Here we shall comment on some additional features of the model.
First, the inclusion of large discrete symmetries might have
an accidental continuous symmetry giving pseudo-Goldstone
bosons. In the present model such flavor invariant operators
that might lead to an accidental symmetry, will be determined
by physics above the cut-off. However, as opposed to the 4D case,
where the masses of the pseudo-Goldstone bosons are suppressed by
ratios of flavon VEV's to the fundamental scale, in our case,
the ratios may not be suppressed since the fundamental scale may
not be far above the cutt-off.

Another issue regarding flavor models with a low scale, is their possible 
contribution to flavor changing processes (FCNC) such as the ones described
in \cite{nima1}. For instance, consider an operator of the type
$\sum_a c_a (Q_a D_a)({\bar{Q}_a} {\bar{D_a}})/\Lambda^3$. Though
this appears to be diagonal in the weak basis, when one goes to
the mass-eigenstate basis, it will induce dangerous FCNC, unless
the scale $\lambda$ is of order  $\simeq 100$ TeV, or there is some reason 
that forces the coefficients $c_a$ to cancel at the one percent level. One
possibility to achieve such cancellations would be to embed the
abelian flavor symmetry into a non-abelian one. However, for
the present model we shall simply assume that physics above the
cut-off is responsible for the cancellation between the coefficients.

 Furthermore, a renormalization group analysis  should also be 
incorporated into the fit. However, since in our model 
the flavon fields have a mass of O($\Lambda$), they do not 
participate in the running and the hierarchies 
are affected only by scale-independent factors of O(1)~\cite{dienes}. 
The running might change the overall
scales, and thus it might be necessary to modify the overall scale of
the Yukawa matrices. This can be easily done by changing either the charge
assignments or by enlarging the flavor group. 
Also, we don't view this model as unique, in fact, it might be possible 
to create a model with a discrete Non-Abelian gauge
symmetry that can be broken sequentially. In this case, 
depending on the scales, the flavon fields
can participate in non-trivial ways through the RGE analysis. 

 On the other hand, once we include the flavon field in the spectrum, it may
be possible to generate Higgs-flavon mixing, which would produce
a ``more flavored'' Higgs profile, which could produce
interesting Higgs signals such as $h\to \tau \mu$ \cite{hixldc}.

Finally, one way to avoid the large flavor charges that appeared
in the 5D model would be to extend the model to six dimensions
(or higher). Namely, since the entries of the Yukawa matrices scale
as $\epsilon^{n \delta}$, while the value of $\epsilon$ can be maintained
of order 0.1, one could lower values for $n$ as $\delta$ increases;
the 5D charges would be reduced by one half by just letting the model
live in six-dimensions.

In summary, we have presented in this paper a new framework in extra
dimensions to address the flavor problem, which is based in the
traditional FN mechanism. We have presented a specific model in 5D 
based on a $Z_{6}$ local symmetry, whose particle content is that 
of the SM plus an additional flavon field. In 5D there is one universal 
extra dimension compactified on an $S^1/Z_2$ orbifold with radius of 
compactification  $R = 1/Mc = 3.33 \,{\rm TeV}^{-1}$, and with
a high energy cutoff of $\Lambda = 10$~TeV which is also the flavor scale.
When the flavon fields acquire vacuum expectation values they generate
Yukawa matrices, which are then used to perform fits to 
the observables in both the quark and charged lepton sectors.
The model successfully accommodates all the the data for the quarks and
charged leptons, while for the neutrino masses our scheme seems adequate to
generate the right textures, however a detailed model, as well
as additional fits, and possibly an improved version of the model 
in six-dimensions, will be left for a future publication \cite{progress}.

\begin{center}
{\bf Acknowledgments}
\end{center}We would like to thank Qaisar Shafi, Dany Marfatia, and Chris 
Carone for
helpful conversations and comments. We also thank Paolo Amore for his
comments and for reading the manuscript. 
A.A.'s work was partially by the Alvarez-Buylla fund of the Universidad de
Colima and by PROMEP (SEP-Mexico).

%\appendix

%%%%%%%%%%%%%%%%%%%%%%%%%%%%%%%%%%%%%%%%%%%%%%%%%%%%
\newpage
%
%%%%%%%%%%%%%%%%%%%%%%%%%%%%%%%%%%%%%%%%%%%%%%%%%%%%%
%  TABLES
%%%%%%%%%%%%%%%%%%%%%%%%%%%%%%%%%%%%%%%%%%%%%%%%%%%%%
\begin{table}
\begin{center}
\begin{tabular}{c|c|c}
\multicolumn{3}{c}{$\epsilon = 0.1$ \,\,\, $\chi^2 = 2.29$ } \\ \hline
$u_1 = +0.96$ & $d_1 = +0.94$ & $l_1 = -1.07$ \\
$u_2 = -1.19$ & $d_2 = -0.89$ & $l_2 = -0.89$ \\
$u_3 = +0.73$ & $d_3 = +1.72$ & $l_3 = +0.81$ \\
$u_4 = -1.07$ & $d_4 = +1.12$ & $l_4 = +1.21$ \\
$u_5 = +1.23$ & $d_5 = +2.28$ & $l_5 = +0.69$ \\
$u_6 = -0.64$ & $d_6 = -0.95$ & $l_6 = +1.22$ \\
$u_7 = -0.98$ & $d_7 = +1.19$ & $l_7 = +1.15$ \\
$u_8 = -0.82$ & $d_8 = -1.97$ & $l_8 = +0.83$ \\
$u_9 = +0.99$ & $d_9 = +1.41$ & $l_9 = +0.58$ \\ \hline
\end{tabular}
\caption{Parameters obtained in one fit for the $Z_{6}$ model.}
\label{fit}
\end{center}
\end{table}

\begin{table}
\begin{center}
\begin{tabular}{lll}
Observable & Expt. value & Fit value \\ \hline
$m_e$ & $( 5.11 \pm 1\% ) \times 10^{-4}$ & $5.11 \times 10^{-4}$ \\
$m_\mu$ & $0.106 \pm 1\%$ & $0.106$ \\
$m_\tau$ & $1.78 \pm 1\%$ & $1.78$ \\
$m_u$ & $( 3.25 \pm 1.8) \times 10^{-3}$ & $3.28\times 10^{-3}$ \\
$m_d$ & $( 6.0 \pm 3.0) \times 10^{-3}$ & $5.69 \times 10^{-3}$ \\
$m_c$ & $1.25 \pm 0.15$ & $1.25$ \\
$m_s$ & $0.115 \pm 0.055$ & $1.04\times 10^{-1}$ \\
$m_t$ & $173.8 \pm 5.2$ & $173.9$ \\
$m_b$ & $4.25 \pm 0.15$ & $4.24$ \\
$|V_{us}|$ & $0.22 \pm 0.0035$ & $0.221$ \\
$|V_{ub}|$& $(3.1 \pm 1.35) \times 10^{-3}$ & $3.1\times 10^{-3}$ \\
$|V_{cb}|$& $0.04 \pm 0.03$ & $0.03$ \\ \hline
\end{tabular}
\caption{Experimental values versus fit central values for observables
using the inputs of Table~\ref{fit}.  
Masses are in GeV and all other
quantities are dimensionless. Error bars are taken as indicated in the text.}
\label{data}
\end{center}
\end{table}
%%%%%%%%%%%%%%%%%%%%%%%%%%%%%%%%%%%%%%%%%%%%%%%%%%%%%%%%%%%%%%%%%

\begin{thebibliography}{99}
\frenchspacing

\bibitem{XD} This is a very partial list: \\
K.~R.~Dienes, E.~Dudas and T.~Gherghetta,
%``Extra spacetime dimensions and unification,''
Phys.\ Lett.\ B {\bf 436}, 55 (1998)
[arXiv:hep-ph/9803466];
%%CITATION = HEP-PH 9803466;%%
%\cite{Arkani-Hamed:1998rs}
%\bibitem{Arkani-Hamed:1998rs}
N.~Arkani-Hamed, S.~Dimopoulos and G.~R.~Dvali,
%``The hierarchy problem and new dimensions at a millimeter,''
Phys.\ Lett.\ B {\bf 429}, 263 (1998)
[arXiv:hep-ph/9803315];
%%CITATION = HEP-PH 9803315;%%
%\cite{Arkani-Hamed:1998nn}
%\bibitem{Arkani-Hamed:1998nn}
N.~Arkani-Hamed, S.~Dimopoulos and G.~R.~Dvali,
%``Phenomenology, astrophysics and cosmology of theories with  sub-millimeter dimensions and TeV scale quantum
gravity,''
Phys.\ Rev.\ D {\bf 59}, 086004 (1999)
[arXiv:hep-ph/9807344];
%%CITATION = HEP-PH 9807344;%%
%\cite{Pomarol:1998sd}
%\bibitem{Pomarol:1998sd}
A.~Pomarol and M.~Quiros,
%``The standard model from extra dimensions,''
Phys.\ Lett.\ B {\bf 438}, 255 (1998)
[arXiv:hep-ph/9806263];
%%CITATION = HEP-PH 9806263;%%
%\cite{Kakushadze:1998vr}
%\bibitem{Kakushadze:1998vr}
Z.~Kakushadze,
%``Novel extension of MSSM and 'TeV scale' coupling unification,''
Nucl.\ Phys.\ B {\bf 548}, 205 (1999)
[arXiv:hep-th/9811193];
%%CITATION = HEP-TH 9811193;%%
%\cite{Carone:1999cb}
%\bibitem{Carone:1999cb}
C.~D.~Carone,
%``Gauge unification in nonminimal models with extra dimensions,''
Phys.\ Lett.\ B {\bf 454}, 70 (1999)
[arXiv:hep-ph/9902407];
%%CITATION = HEP-PH 9902407;%%
%\cite{Carone:1999nz}
%\bibitem{Carone:1999nz}
C.~D.~Carone,
%``Electroweak constraints on extended models with extra dimensions,''
Phys.\ Rev.\ D {\bf 61}, 015008 (2000)
[arXiv:hep-ph/9907362];
%%CITATION = HEP-PH 9907362;%%
%\cite{Huber:2001gw}
%\bibitem{Huber:2001gw}
S.~J.~Huber, C.~A.~Lee and Q.~Shafi,
%``Kaluza-Klein excitations of W and Z at the LHC?,''
Phys.\ Lett.\ B {\bf 531}, 112 (2002)
[arXiv:hep-ph/0111465].
%%CITATION = HEP-PH 0111465;%%
\bibitem{randall}%\cite{Randall:1999vf}
%\bibitem{Randall:1999vf}
L.~Randall and R.~Sundrum,
%``An alternative to compactification,''
Phys.\ Rev.\ Lett.\  {\bf 83}, 4690 (1999)
[arXiv:hep-th/9906064];
%%CITATION = HEP-TH 9906064;%%
%\cite{Randall:1999ee}
%\bibitem{Randall:1999ee}
L.~Randall and R.~Sundrum,
%``A large mass hierarchy from a small extra dimension,''
Phys.\ Rev.\ Lett.\  {\bf 83}, 3370 (1999)
[arXiv:hep-ph/9905221].
%%CITATION = HEP-PH 9905221;%%
\bibitem{flavor}%\cite{Huber:2002gp}
%\bibitem{Huber:2002gp}
S.~J.~Huber and Q.~Shafi,
%``Majorana neutrinos in a warped 5D standard model,''
arXiv:hep-ph/0205327;
%%CITATION = HEP-PH 0205327;%%
%\cite{Huber:2000ie}
%\bibitem{Huber:2000ie}
S.~J.~Huber and Q.~Shafi,
%``Fermion masses, mixings and proton decay in a Randall-Sundrum model,''
Phys.\ Lett.\ B {\bf 498}, 256 (2001)
[arXiv:hep-ph/0010195];
%%CITATION = HEP-PH 0010195;%%
%\cite{Grossman:1999ra}
%\bibitem{Grossman:1999ra}
Y.~Grossman and M.~Neubert,
%``Neutrino masses and mixings in non-factorizable geometry,''
Phys.\ Lett.\ B {\bf 474}, 361 (2000)
[arXiv:hep-ph/9912408];
%%CITATION = HEP-PH 9912408;%%
%\cite{Dooling:2002js}
%\bibitem{Dooling:2002js}
D.~Dooling, D.~A.~Easson and K.~Kang,
%``Geometric origin of CP violation in an extra-dimensional brane world,''
arXiv:hep-ph/0202206;
%%CITATION = HEP-PH 0202206;%%
%\cite{Appelquist:2002ft}
%\bibitem{Appelquist:2002ft}
T.~Appelquist, B.~A.~Dobrescu, E.~Ponton and H.~U.~Yee,
%``Neutrinos vis-a-vis the six-dimensional standard model,''
Phys.\ Rev.\ D {\bf 65}, 105019 (2002)
[arXiv:hep-ph/0201131].
%%CITATION = HEP-PH 0201131;%%
%%CITATION = HEP-PH 0006278;%%
\bibitem{nima1}%\cite{Arkani-Hamed:1999yy}
%\bibitem{Arkani-Hamed:1999yy}
N.~Arkani-Hamed, L.~J.~Hall, D.~R.~Smith and N.~Weiner,
%``Flavor at the TeV scale with extra dimensions,''
Phys.\ Rev.\ D {\bf 61}, 116003 (2000)
[arXiv:hep-ph/9909326].
%%CITATION = HEP-PH 9909326;%%
\bibitem{martin1}%\cite{Mirabelli:1999ks}
%\bibitem{Mirabelli:1999ks}
E.~A.~Mirabelli and M.~Schmaltz,
%``Yukawa hierarchies from split fermions in extra dimensions,''
Phys.\ Rev.\ D {\bf 61} (2000) 113011
[arXiv:hep-ph/9912265].
%%CITATION = HEP-PH 9912265;%%
\bibitem{fat}%\cite{Georgi:2000wb}
%\bibitem{Georgi:2000wb}
H.~Georgi, A.~K.~Grant and G.~Hailu,
%``Chiral fermions, orbifolds, scalars and fat branes,''
Phys.\ Rev.\ D {\bf 63}, 064027 (2001)
[arXiv:hep-ph/0007350];
%%CITATION = HEP-PH 0007350;%%
%\cite{Haba:2002uw}
%\bibitem{Haba:2002uw}
N.~Haba and N.~Maru,
%``(S)fermion masses in fat brane scenario,''
arXiv:hep-ph/0204069.
%%CITATION = HEP-PH 0204069;%%
\bibitem{dvali1}%\cite{Dvali:2000ha}
%\bibitem{Dvali:2000ha}
G.~R.~Dvali and M.~A.~Shifman,
%``Families as neighbors in extra dimension,''
Phys.\ Lett.\ B {\bf 475}, 295 (2000)
[arXiv:hep-ph/0001072];
%%CITATION = HEP-PH 0001072;%%
%\cite{Kaplan:2001ga}
%\bibitem{Kaplan:2001ga}
D.~E.~Kaplan and T.~M.~Tait,
%``New tools for fermion masses from extra dimensions,''
JHEP {\bf 0111}, 051 (2001)
[arXiv:hep-ph/0110126].
%%CITATION = HEP-PH 0110126;%%
%\cite{Matsuda:2002vf}
%\bibitem{Matsuda:2002vf}
T.~Matsuda,
%``Activated sphalerons and large extra dimensions,''
arXiv:hep-ph/0205331.
%%CITATION = HEP-PH 0205331;%%
\bibitem{dobrescu1}%\cite{Appelquist:2000nn}
%\bibitem{Appelquist:2000nn}
T.~Appelquist, H.~C.~Cheng and B.~A.~Dobrescu,
%``Bounds on universal extra dimensions,''
Phys.\ Rev.\ D {\bf 64}, 035002 (2001)
[arXiv:hep-ph/0012100].
%%CITATION = HEP-PH 0012100;%%
\bibitem{green}
%\cite{Green:sg}
%\bibitem{Green:sg}
M.~B.~Green and J.~H.~Schwarz,
%``Anomaly Cancellation In Supersymmetric D=10 Gauge Theory And Superstring Theory,''
Phys.\ Lett.\ B {\bf 149}, 117 (1984).
%%CITATION = PHLTA,B149,117;%%
\bibitem{discrete}
J. Preskill, in {\it Architecture of Fundamental Interactions at Short
Distances}, Proceedings of the Les Houches Summer School of Theoretical
Physics, Session XLIV, 1985, edited by P. Ramond and R. Stora 
(Elsevier, New York, 1987), p. 235;
%\cite{Aranda:2000tm}
%\bibitem{Aranda:2000tm}
A.~Aranda, C.~D.~Carone and R.~F.~Lebed,
%``Maximal neutrino mixing from a minimal flavor symmetry,''
Phys.\ Rev.\ D {\bf 62}, 016009 (2000)
[arXiv:hep-ph/0002044].
%%CITATION = HEP-PH 0002044;%%
\bibitem{aranda}%\cite{Aranda:2001rd}
%\bibitem{Aranda:2001rd}
A.~Aranda, C.~D.~Carone and P.~Meade,
%``U(2)-like flavor symmetries and approximate bimaximal neutrino mixing,''
Phys.\ Rev.\ D {\bf 65}, 013011 (2002)
[arXiv:hep-ph/0109120].
%%CITATION = HEP-PH 0109120;%%
\bibitem{PDG}
%\cite{Groom:in}
%\bibitem{Groom:in}
D.~E.~Groom {\it et al.}  [Particle Data Group Collaboration],
%``Review Of Particle Physics,''
Eur.\ Phys.\ J.\ C {\bf 15}, 1 (2000).
%%CITATION = EPHJA,C15,1;%%
\bibitem{superk}%\cite{:2002pe}
%\bibitem{:2002pe}
[Super-Kamiokande Collaboration],
%``Determination of solar neutrino oscillation parameters using 1496 days  of Super-Kamiokande-I data,''
arXiv:hep-ex/0205075.
%%CITATION = HEP-EX 0205075;%%
\bibitem{atmos}%\cite{Kajita:zv}
%\bibitem{Kajita:zv}
T.~Kajita  [Super-Kamiokande Collaboration],
%``Study Of Neutrino Oscillations With The Atmospheric Neutrino Data From Superkamiokande,''
Nucl.\ Phys.\ Proc.\ Suppl.\  {\bf 100}, 107 (2001).
%%CITATION = NUPHZ,100,107;%%
\bibitem{progress} A. Aranda and L. Diaz-Cruz, Work in progress.
\bibitem{13}
%\cite{King:2001uq}
%\bibitem{King:2001uq}
S.~F.~King,
%``Neutrino oscillations: Status, prospects and opportunities at a  neutrino factory,''
J.\ Phys.\ G {\bf 27}, 2149 (2001)
[arXiv:hep-ph/0105261].
%%CITATION = HEP-PH 0105261;%%
\bibitem{cosmo}
%\cite{Elgaroy:2002bi}
%\bibitem{Elgaroy:2002bi}
O.~Elgaroy {\it et al.},
%``A new limit on the total neutrino mass from the 2dF galaxy redshift  survey,''
arXiv:astro-ph/0204152.
%%CITATION = ASTRO-PH 0204152;%%
\bibitem{LEP}
%\cite{Abbaneo:2001ix}
%\bibitem{Abbaneo:2001ix}
D.~Abbaneo {\it et al.}  [ALEPH Collaboration],
%``A combination of preliminary electroweak measurements and constraints  on the standard model,''
arXiv:hep-ex/0112021.
%%CITATION = HEP-EX 0112021;%%
\bibitem{dienes}%\cite{Dienes:1998vg}
%\bibitem{Dienes:1998vg}
K.~R.~Dienes, E.~Dudas and T.~Gherghetta,
%``Grand unification at intermediate mass scales through extra dimensions,''
Nucl.\ Phys.\ B {\bf 537}, 47 (1999)
[arXiv:hep-ph/9806292].
%%CITATION = HEP-PH 9806292;%%

\bibitem{hixldc} J.~L.~Diaz-Cruz,
%``A more flavored Higgs boson in supersymmetric models,''
JHEP {\bf 0305}, 036 (2003) [arXiv:hep-ph/0207030];
%%CITATION = HEP-PH 0207030;%%
J.~L.~Diaz-Cruz and J.~J.~Toscano,
 %``Probing lepton flavour violation with the Higgs boson decays H $\to$  l/i+
%l/j-,''
Phys.\ Rev.\ D {\bf 62}, 116005 (2000)
[arXiv:hep-ph/9910233];
%%CITATION = HEP-PH 9910233;%%
J.~L.~Diaz-Cruz, R.~Noriega-Papaqui and A.~Rosado,
%``Mass matrix Ansatz and lepton flavor violation in the THDM-III,''
arXiv:hep-ph/0401194.
%%CITATION = HEP-PH 0401194;%% 
\end{thebibliography}
\end{document}